\documentclass[a4paper,11pt]{article}
\pdfoutput=1
\usepackage{jheppub}
\usepackage{amsmath}
\usepackage{amssymb}
\usepackage{color}
\usepackage{xcolor}
\usepackage{graphicx}
\usepackage{hyperref}
\usepackage{xspace,slashed}
\usepackage[utf8]{inputenc}
\usepackage{subcaption}
\usepackage{multirow}
\usepackage{nccmath} 

\makeatletter
\renewcommand\@fpheader{}
\renewcommand\@journal{}
\makeatother

\newcommand{\as}{\alpha_s}
\newcommand{\ep}{\epsilon}
\newcommand{\oT}{\overline{T}}
\newcommand{\oF}{\overline{\mathcal{F}}}
\newcommand{\lambdavec}{{\vec{\lambda}}}
\DeclareMathOperator{\Li}{Li}

\newcommand{\OXaff}{Rudolf Peierls Centre for Theoretical Physics,
  Clarendon Laboratory, Parks Road, Oxford OX1 3PU, UK}
\newcommand{\WADaff}{Wadham College, Oxford OX1 3PN, UK}
\newcommand{\TUMaff}{Physik Department, 
  Technische Universit\"at M\"unchen, James-Franck-Stra\ss{}e 1, 85748 Garching, Germany}
\newcommand{\MSUaff}{Department  of  Physics  and  Astronomy,  Michigan  State  University, East  Lansing,  Michigan  48824,  USA}

\preprint{
    OUTP-21-27P,
    MSUHEP-21-032,
    TUM-HEP-1372/21
}

\title{
  Three-loop helicity amplitudes for diphoton production in gluon fusion
}
\author[a]{Piotr Bargie\l{}a,}
\author[a,b]{Fabrizio Caola,}
\author[c]{Andreas von Manteuffel,}
\author[a,d]{Lorenzo Tancredi}
\affiliation[a]{\OXaff}
\affiliation[b]{\WADaff}
\affiliation[c]{\MSUaff}
\affiliation[d]{\TUMaff}

\abstract{
   We present a calculation of the helicity amplitudes for the process
   $gg\to\gamma\gamma$ in three-loop massless QCD. We employ a recently
  proposed method to calculate scattering amplitudes in the
  't Hooft-Veltman scheme that reduces the amount of
  spurious non-physical information needed at intermediate stages of
  the computation. Our analytic results for the three-loop helicity amplitudes
  are remarkably compact, and can be efficiently evaluated numerically.
  This calculation provides the last missing building block for the computation
  of NNLO QCD corrections to diphoton production in gluon fusion.}

\begin{document}

\maketitle 
\section{Introduction}
\label{sec:intro}
The production of two hard photons
is an important process at 
hadron colliders, which both allows for scrutiny of the structure of the
Standard Model and serves as an important background for many Higgs
and new physics analyses.

From a theoretical perspective, the $pp\to\gamma\gamma$ process is rather
peculiar. Phenomenologically, this process is 
interesting because an operative definition of isolated photons is
non-trivial, and it requires quite subtle theoretical
analysis~\cite{Frixione:1998jh}.  Computationally,
diphoton production is relatively simple, yet non-trivial. Indeed,
photons are massless and colour-neutral particles, which implies that
both the infrared structure and the scattering amplitudes for the
diphoton process are not very complicated. However, compared to other
colour-singlet processes like Higgs or Drell-Yan production, the kinematics of $\gamma\gamma$ production is more
involved as it depends non-trivially on a scattering angle already at
leading order (LO) in the perturbative expansion. Because of
these features, diphoton production is an ideal process for
testing and improving our understanding of Quantum Chromodynamics (QCD) at hadron colliders.
Indeed, historically $\gamma\gamma$ production has often served as a
testing ground for innovative studies in perturbative QCD.  For
example, $\gamma\gamma$ production was the
first hadron collider process with non-trivial LO kinematics for which
next-to-next-to-leading order (NNLO) QCD corrections were
computed~\cite{Catani:2011qz}. Furthermore, $q\bar q\to\gamma\gamma$ was the
first $2\to2$ QCD scattering amplitude that was calculated 
at the three-loop level~\cite{Caola:2021rqz}. Photon processes also played a
prominent role in the development of NNLO predictions for
$2\to 3$ collider reactions~\cite{Chawdhry:2019bji,
  Chawdhry:2020for,Chawdhry:2021mkw,Chawdhry:2021hkp,Agarwal:2021grm,Agarwal:2021vdh,
  Badger:2021imn,Badger:2021ohm}. 

The leading mechanism for producing two photons at hadron colliders
is through $q\bar q$ annihilation.
The availability of the two-loop QCD scattering amplitudes
for $q\bar q\to\gamma\gamma$~\cite{Anastasiou:2002zn} enabled
detailed phenomenological predictions at NNLO accuracy \cite{Catani:2011qz,Campbell:2016yrh,Gehrmann:2020oec,
  Grazzini:2017mhc,Alioli:2020qrd,Neumann:2021zkb}.
Starting from
NNLO, the $gg$ partonic channel opens up. There
are two contributions to this: tree-level corrections of the
form $gg\to \gamma\gamma + q\bar q$ and loop-induced corrections
$gg\to\gamma\gamma$. Phenomenologically, the former are 
typically very small and we will not discuss them further. 
The loop-induced contribution is instead quite interesting. 
 First, the large gluon flux at the Large Hadron Collider (LHC) compensates for
the $\as$ suppression, making it
important for precision phenomenological studies.  Being a new
channel, it has all the features of a leading order process, in particular
large perturbative uncertainties. Moreover, being gluon induced one
expects particularly large radiative corrections. This has spurred many
investigations, which upgraded the precision in this channel to
next-to-leading order (NLO) in QCD, i.e.\ to $\mathcal
O(\as^3)$~\cite{Bern:2001df,Bern:2002jx}.

Given the ever-increasing
experimental precision on diphoton measurements~\cite{CMS:2014mvm,ATLAS:2021mbt},
it becomes interesting to try and push the theoretical precision even
further and consider NNLO corrections to the $gg\to\gamma\gamma$
process. While this is desirable for a variety of LHC analyses, it is
of particular importance for Higgs studies. Indeed, in this case there
is a subtle signal/background interference effect between the $gg\to
H\to\gamma\gamma$ signal and the continuum $gg\to\gamma\gamma$
background, which is known to modify the Higgs
line-shape~\cite{Martin:2012xc}. This effect can in turn be used to
constrain the Higgs boson total decay width~\cite{Dixon:2013haa}.
This kind of investigations require an exquisite experimental control,
see e.g. ref.~\cite{LHCHiggsCrossSectionWorkingGroup:2016ypw}, as well
as robust control of theoretical predictions for both the signal and
the background processes. Several in-depth
analysis~\cite{Martin:2013ula,deFlorian:2013psa,Coradeschi:2015tna,Campbell:2017rke}
suggest that reaching NNLO QCD accuracy in the gluon channel is
desirable.
A major step towards the calculation of full NNLO QCD corrections to
$gg\to\gamma\gamma$ has been made very recently with the computation of NLO
QCD corrections to the $gg\to\gamma\gamma+j$
process~\cite{Badger:2021imn,Badger:2021ohm}. In this paper, we present
a calculation of the last missing ingredient, 
the three loop virtual amplitude for the $gg\to\gamma\gamma$ process.

The rest of this paper is organised as follows. In sec.~\ref{sec:not}
we set up our notation and discuss the generic kinematics features of
the $gg\to\gamma\gamma$ process. In sec.~\ref{sec:hel} we briefly
review the approach of refs~\cite{Peraro:2019cjj,Peraro:2020sfm} to
the calculation of helicity amplitudes that we adopt here. In
sec.~\ref{sec:comp} we provide more technical details on our three-loop
calculation. In sec.~\ref{sec:uvir} we discuss the ultraviolet and
infrared structure of the scattering amplitude, and define the renormalised
finite remainders which are the main result of this paper. In
sec.~\ref{subsec:res} we document the checks that we have performed
on our calculation, and briefly describe the general structure of our
result. We also present analytic formulas for the three loop finite
remainder for the simplest helicity configuration. The analytic formulas
for all the relevant helicity configurations can be found in computer-readable
format in the ancillary material that accompany this submission.
Finally, we conclude in sec.~\ref{sec:concl}.

\section{Notation and kinematics}
\label{sec:not}

We consider virtual QCD corrections to the production of two
photons through gluon fusion
\begin{equation}
g (p_1) + g(p_2)  \to \gamma(-p_3) + \gamma(-p_4)\,,
\label{eq:ggaa}
\end{equation}
mediated by light quarks.
The signs of the momenta are chosen such that all momenta are incoming,
$p_1+p_2+p_3+p_4=0$.
All particles in the process are on the mass-shell, $p_1^2=p_2^2=p_3^2=p_4^2=0$.
The kinematics is fully described by 
the usual Mandelstam invariants
\begin{equation}\label{eq:physregion}
s = (p_1+ p_2)^2\,,~ t = (p_1+p_3)^2\,,~u = (p_2+p_3)^2 \,\,, \quad s+t+u = 0\,.
\end{equation}
In the physical scattering region, one has $s>0$, $t<0$, $u<0$.
For later reference, we also introduce the dimensionless ratio
\begin{equation}
    x = -\frac{t}{s},
\end{equation}
where $0 < x < 1$ in the physical region.
We work in $d=4-2\ep$ dimensions to regulate ultraviolet (UV) and infrared (IR) divergences. To
be precise, we adopt the 't Hooft-Veltman (tHV)
scheme~\cite{THOOFT1972189}, i.e. we perform computations for generic
$d$ but we constrain all the external particles and their polarisations to
live in the physical $d=4$ subspace.  
This allows us to simplify
the calculation compared to the Conventional Dimensional Regularisation
(CDR) case, where internal and external degrees of freedom are treated
as $d$-dimensional.

We write the scattering amplitude for the process in eq.~(\ref{eq:ggaa}) as
\begin{align}
\mathcal{A}(s,t) &= 
 \delta^{a_1 a_2} (4 \pi \alpha) A(s,t) \nonumber \\
&= \delta^{a_1 a_2} (4 \pi \alpha) A^{\mu \nu \rho
  \sigma}(s,t) \epsilon_{1,\mu}(p_1) \epsilon_{2,\nu}(p_2)
\epsilon_{3,\rho}(p_3) \epsilon_{4,\sigma}(p_4),
\label{eq:amp}
\end{align}
where $a_j$ is the colour index of the gluon of momentum $p_j$ and
$\epsilon_{j,\mu}(p_j)$ is the polarisation vector of the 
vector boson of momentum $p_j$. For convenience, we have extracted
the leading-order electroweak coupling written in terms of the fine
structure constant $\alpha$, where $e = \sqrt{4 \pi \alpha}$ is the
unit of electric charge. We are interested in the QCD perturbative expansion of
eq.~(\ref{eq:amp})
\begin{equation}
 A(s,t) =
  \frac{\alpha_s}{2\pi} \left[ A^{(1)}(s,t) + \frac{\alpha_s}{2\pi}
  A^{(2)}(s,t) + 
  \left(\frac{\alpha_s}{2\pi}\right)^2
   A^{(3)}(s,t) + \mathcal O(\alpha_s^3)\right],
\end{equation}
where $\alpha_s=\alpha_s(\mu)$ is the $\overline{\rm MS}$ renormalized
QCD coupling and the superscript indicates the number of loops $L$.
We find it convenient to express the result for $A^{(L)}$
in terms of the quadratic Casimir invariants
of theory $C_A$ and $C_F$. They are defined through
\begin{equation}
T^a_{ij}T^a_{jk} = C_F \delta_{ik} \,,\qquad 
f^{acd}f^{bcd} = C_A \delta^{a b} \,,
\end{equation}
where $f^{abc}$ and $T^{a}_{ij}$ are the $SU(3)$ structure constants and the generators
in the fundamental representation, respectively. We normalise the generators as
\begin{equation}
  {\rm Tr}[T^a T^b] = T_F \delta^{a_1 a_2},~~~~ T_F = \frac{1}{2}.
\end{equation}
In QCD, $C_A=3$ and $C_F=4/3$.

The Feynman diagrams for the process eq.~(\ref{eq:ggaa}) can be naturally separated
according to whether the two photons couple to the same or to two different closed
fermion lines. We then introduce the following short hands for the respective electromagnetic coupling structures
\begin{equation}
  \big(n_f^V\big)^2 = \big(\sum_f Q_f\big)^2,
  \quad
  n_f^{V_2}=\sum_f Q_f^2 \,,
  \label{eq:nfdef}
\end{equation}
where the sums run over $n_f$ light quarks
and $Q_f$ is their charge in units of $e$, 
i.e.\ $Q_{u,c} = 2/3$, $Q_{d,s,b} = -1/3$. For QCD with 5 flavours, the
structures in eq.~(\ref{eq:nfdef}) evaluate to $(n_f^V)^2 = (1/3)^2 = 1/9$
and $n_f^{V_2} = 11/9$. 

\section{The helicity amplitudes}
\label{sec:hel}
In this section, we explain how one can efficiently calculate the amplitude in eq.~(\ref{eq:amp}) for specific helicities. We start by discussing the tensor
$A^{\mu\nu\rho\sigma}$. It can be expanded as
\begin{equation}
  A^{\mu\nu\rho\sigma}(s,t) =
  \sum_i \mathcal F_i(s,t) \Gamma_i^{\mu\nu\rho\sigma},
\end{equation}
where $\mathcal F_i$ are scalar form factors\footnote{We note that the
  form factors $\mathcal F_i$ also depend on the dimension of the
  space-time. This dependence is assumed.} and
$\Gamma^{\mu\nu\rho\sigma}_i$ are independent tensor structures
constructed using external momenta $\{p_i^\mu\}$ and the metric tensor
$g^{\mu\nu}$.  With three independent external momenta, the total
number of tensor structures that one can write is 138, see
e.g.~\cite{Binoth:2002xg}. Since $A^{\mu\nu\rho\sigma}$ has to be
contracted with the external polarisation vectors $\epsilon_i^\mu$,
one can use the physical conditions $p_i\cdot \epsilon_i = 0$ to
remove all tensors proportional to $p_1^\mu$, $p_2^\nu$, $p_3^\rho$,
$p_4^\sigma$. This removes all but 57 structures. By making a specific
choice for the reference vectors of the external gauge bosons, one may
eliminate further redundancies.  A convenient choice is to impose
\begin{align}
\epsilon_i \cdot p_{i+1} = 0 \,,\;\; \mbox{where} \;\; i=1,...,4 \;\; \mbox{and} \;\; p_5\equiv p_1.
\label{eq:gauge}
\end{align}
This leaves one with 10 independent structures, that we choose as
\begin{align} \label{eq:structs}
\Gamma_1^{\mu \nu \rho \sigma} &= p_3^{\mu}p_1^{\nu}p_1^{\rho}p_2^{\sigma}\,, \;\;
\Gamma_2^{\mu \nu \rho \sigma} = p_3^{\mu}p_1^{\nu}g^{\rho\sigma}\,, \nonumber \\
\Gamma_3^{\mu \nu \rho \sigma} &= p_3^{\mu}p_1^{\rho}g^{\nu\sigma} \,\,\, \,, \;\;
\Gamma_4^{\mu \nu \rho \sigma} = p_3^{\mu}p_2^{\sigma}g^{\nu\rho}\,, \nonumber \\
\Gamma_5^{\mu \nu \rho \sigma} &= p_1^{\nu}p_1^{\rho}g^{\mu\sigma} \,\,\, \,, \;\;
\Gamma_6^{\mu \nu \rho \sigma} = p_1^{\nu}p_2^{\sigma}g^{\mu\rho}\,, \nonumber \\
\Gamma_7^{\mu \nu \rho \sigma} &= p_1^{\rho}p_2^{\sigma}g^{\mu\nu} \,\,\, \,, \;\;
\Gamma_8^{\mu \nu \rho \sigma} = g^{\mu\nu}g^{\rho\sigma}\,, \nonumber \\
\Gamma_9^{\mu \nu \rho \sigma} &= g^{\mu\sigma}g^{\nu\rho} \,\,\,\,\,\, \,, \;\;
\Gamma_{10}^{\mu \nu \rho \sigma} = g^{\mu\rho}g^{\nu\sigma}.
\end{align}
For notational convenience, we define the 10 independent structures
\begin{equation}
T_i =  \Gamma_i^{\mu \nu \rho \sigma}\, \epsilon_{1,\mu} \epsilon_{2,\nu} \epsilon_{3,\rho} \epsilon_{4,\sigma}
\label{eq:tensors}
\end{equation}
and refer to them, with a slight abuse of language, as \emph{tensors}.
The scattering amplitude eq.~(\ref{eq:amp}) can then be written as
\begin{equation}
  A(s,t) = 
  \sum_{i=1}^{10} \mathcal F_i(s,t) T_i.
  \label{eq:ampT}
\end{equation}
We stress that eq.~(\ref{eq:ampT}) is valid at any perturbative order and for any space-time
dimension.

In four dimensions, it is easy to see that only 8 out of the 10
tensors $T_i$ are actually independent. 
It turns out that in the tHV scheme, it is possible to separate the purely
four-dimensional tensor structures from the $-2\ep$-dimensional ones through a simple
orthogonalisation procedure~\cite{Peraro:2019cjj,Peraro:2020sfm}.  
We briefly sketch how this can be done for our process, and refer
the reader to refs~\cite{Peraro:2019cjj,Peraro:2020sfm} for a
thorough discussion. Following ref.~\cite{Peraro:2020sfm}, 
we introduce a new tensor basis $\oT_i$
\begin{equation}
  A(s,t) = 
  \sum_{i=1}^{10} \oF_i(s,t) \oT_i\,,
  \label{eq:ampoT}  
\end{equation}
where the first 7 tensors are identical to the ones introduced before
\begin{equation}\label{eq:tens17}
\oT_i = T_i\,, \quad  i=1,...,7,
\end{equation}
while $\oT_8$ is a symmetrised version of $T_8$
\begin{equation}
\oT_8 = T_8 + T_9 +T_{10}.
\end{equation}
It turns out~\cite{Peraro:2020sfm} that these 8 tensors span the
physical $d=4$ subspace and do not have any component in the
$-2\ep$ directions. The last two
tensors $\oT_{9,10}$ can then be chosen in such a way that they
are constrained to live in the $-2\ep$ subspace. 
This can be achieved  by simply removing from
the original $T_{9,10}$ their projection along $\oT_{1...8}$
\begin{equation}
\label{eq:tens2}
\oT_i = T_i - \sum_{j=1}^{8}(\mathcal{P}_j T_i) \oT_j \,,~ i=9,10 \,,
\end{equation}
where the projectors $\mathcal P_i$ are defined through
\begin{equation}
  \sum_{\rm pol} \mathcal P_i \oT_j = \delta_{ij}.
\end{equation}
The explicit form of the $\mathcal P_i$ projectors relevant for our case
can be found in ref.~\cite{Peraro:2020sfm}.
The new tensors $\oT_{9,10}$ read
\begin{equation}
\begin{split}
& \oT_9 \,\, = {T_{9}} \,\,\, - \frac{1}{3}\left( -\frac{2 {\oT_1}}{s u} - \,\,\, \frac{{\oT_6}}{s} - \frac{{\oT_2}+{\oT_3}+2 {\oT_4}-2 {\oT_5}-{\oT_6}-{\oT_7}}{t} + \,\,\frac{{\oT_3}}{u} + {\oT_8} \right) \,, \\
&\oT_{10} = {T_{10}} - \frac{1}{3} \left( \,\,\,\,\, \frac{4 {\oT_1}}{s u} + \frac{2 {\oT_6}}{s} - \frac{{\oT_2}-2 {\oT_3}-{\oT_4}+{\oT_5}+2 {\oT_6}-{\oT_7}}{t} - \frac{2 {\oT_3}}{u} + {\oT_8} \right) \,.
\label{eq:barT}
\end{split}
\end{equation}\

The  tensors $\oT_{9,10}$ so constructed identically
vanish if they are computed using physical  polarisation vectors in $d=4$ space-time dimensions
and can be safely dropped if one is after tHV helicity
amplitudes~\cite{Peraro:2020sfm}. 
For a given helicity configuration we then write
\begin{equation}
  A_{\lambda_1\lambda_2\lambda_3\lambda_4}(s,t)
  = \sum_{i=1}^{8} \oF_i(s,t)
  \oT_{i,\lambda_1\lambda_2\lambda_3\lambda_4},
\end{equation}
where $\oT_{i,\lambda_1\lambda_2\lambda_3\lambda_4}$ are the tensors
evaluated with polarisation vectors for well-defined helicity states
$\lambda_i$. It should not be surprising that the generic helicity
amplitude can be parametrised in terms of 8 independent structures.
Indeed, in four dimensions we would need to consider $2^4=16$ independent
helicity amplitudes. However, half of them can be related by parity,
which leaves us with 8 independent helicity states. These are in one-to-one
correspondence with the 8 form factors $\oF_i$.

When dealing with helicity amplitudes, we find it convenient to factor
out a spinor function carrying the relevant helicity weight. We achieve this by writing
\begin{equation}
  A_{\lambda_1\lambda_2\lambda_3\lambda_4}(s,t)
  = \mathcal S_{\lambda_1\lambda_2\lambda_3\lambda_4}\, 
  f_{\lambda_1\lambda_2\lambda_3\lambda_4}(s,t),
\end{equation}
where 
\begin{align}
 \mathcal{S}_{++++} &= \frac{[1 2][3 4]}{\langle1 2\rangle\langle3
    4\rangle} \,, & 
    \mathcal{S}_{-+++} &=
  \frac{\langle1 2\rangle\langle1 4\rangle[2 4]}{\langle3
    4\rangle\langle2 3\rangle\langle2 4\rangle} \,, &
  \mathcal{S}_{+-++} &= \frac{\langle2 1\rangle\langle2 4\rangle[1
      4]}{\langle3 4\rangle\langle1 3\rangle\langle1 4\rangle} \,,
  \notag\\
  \mathcal{S}_{++-+} &= \frac{\langle3 2\rangle\langle3 4\rangle[2
      4]}{\langle1 4\rangle\langle2 1\rangle\langle2 4\rangle} \,,&
  \mathcal{S}_{+++-} &= \frac{\langle4 2\rangle\langle4 3\rangle[2
      3]}{\langle1 3\rangle\langle2 1\rangle\langle2 3\rangle} \,,&
   \mathcal{S}_{--++} &= \frac{\langle1 2\rangle[3 4]}{[1
      2]\langle3 4\rangle} \,,
   \notag\\ 
  \mathcal{S}_{-+-+} &= \frac{\langle1 3\rangle[2 4]}{[1 3]\langle2
    4\rangle} \,,&
    \mathcal{S}_{+--+} &= \frac{\langle2 3\rangle[1 4]}{[2 3]\langle1 4\rangle} \,,&&
\label{eq:helamp}
\end{align}
and
\begin{align}
 f_{++++} &=  \frac{t^2}{4}\left(\frac{2\oF_{6}}{u}-\frac{2\oF_{3}}{s}-\oF_{1}\right)+\oF_{8}\left(\frac{s}{u}+\frac{u}{s}+4\right)+\frac{t}{2}(\oF_{2}-\oF_{4}+\oF_{5}-\oF_{7})\,, \notag\\ 
 f_{-+++} &=  \,\,\,\, \frac{t^2}{4}\left(\frac{2\oF_{3}}{s}+\oF_{1}\right)+t\left(\frac{\oF_{8}}{s}+\frac{1}{2}(\oF_{4}+\oF_{6}-\oF_{2})\right)\,, \notag\\ 
 f_{+-++} &=  -\frac{t^2}{4}\left(\frac{2\oF_{6}}{u}-\oF_{1}\right)+t\left(\frac{\oF_{8}}{u}-\frac{1}{2}(\oF_{2}+\oF_{3}+\oF_{5})\right)\,, \notag\\ 
 f_{++-+} &= \,\,\,\, \frac{t^2}{4}\left(\frac{2\oF_{3}}{s}+\oF_{1}\right)+t\left(\frac{\oF_{8}}{s}+\frac{1}{2}(\oF_{6}+\oF_{7}-\oF_{5})\right)\,, \notag\\ 
 f_{+++-} &=  -\frac{t^2}{4}\left(\frac{2\oF_{6}}{u}-\oF_{1}\right)+t\left(\frac{\oF_{8}}{u}+\frac{1}{2}(\oF_{4}+\oF_{7}-\oF_{3})\right)\,, \notag\\ 
 f_{--++} &=  -\frac{t^2}{4}\oF_{1}+\frac{1}{2}t(\oF_{2}+\oF_{3}-\oF_{6}-\oF_{7})+2\oF_{8}\,, \notag\\ 
 f_{-+-+} &=  t^2\left(\frac{\oF_{8}}{su}-\frac{\oF_{3}}{2s}+\frac{\oF_{6}}{2u}-\frac{\oF_{1}}{4}\right)\,, \notag\\ 
 f_{+--+} &=  -\frac{t^2}{4}\oF_{1}+\frac{1}{2}t(\oF_{3}-\oF_{4}+\oF_{5}-\oF_{6})+2\oF_{8}
\,.
\label{eq:alphabeta}
\end{align}
We note that we have chosen the spinor functions in eq.~(\ref{eq:helamp})
following ref.~\cite{Bern:2001df}. The expressions for the spinor-free amplitudes $f_{\lambda_1\lambda_2\lambda_3\lambda_4}$ can be
easily obtained by computing the relevant $\oT_i$ with polarisation
vectors for fixed helicity states. We also note that we define
``$\pm$'' helicity states as\footnote{See e.g.\ ref.~\cite{Dixon:1996wi} for
a review of the spinor-helicity formalism. We follow the notation of ~\cite{Dixon:1996wi}, with the identification $|i^+\rangle = |i\rangle \,, |i^-\rangle = |i] \,, \langle i^+| = [i| \,, \langle i^-| = \langle i|\,,$ and complex conjugation $\langle ij \rangle^* = [ji]\,$.}
\begin{equation}
\epsilon^\mu_{j,-}(p_j) = \frac{\langle p_j | \gamma^\mu | q_j ] }{ \sqrt{2} [ p_j q_j ]}\,, \quad
\epsilon^\mu_{j,+}(p_j) = \frac{\langle q_j | \gamma^\mu | p_j ] }{ \sqrt{2} \langle q_j p_j \rangle }\,,  
\end{equation}
where $q_j$ is the reference vector for the boson $j$, irrespective of whether the particles are in the initial or the final
state.

We have written eqs~(\ref{eq:helamp},\ref{eq:alphabeta}) for only 8 helicity
states. The 8 remaining ones can be obtained from these by exploiting parity
invariance,
\begin{equation}
A_{\lambda_1 \lambda_2 \lambda_3 \lambda_4} =
A_{-\lambda_1,-\lambda_2,-\lambda_3,-\lambda_4}\left(
\langle ij \rangle \leftrightarrow [ji] \right)\,,
\end{equation}
where $-\lambda_i$ indicates the opposite helicity of $\lambda_i$.
We also note that the helicity amplitudes must obey Bose symmetry, i.e.\ 
they must be symmetric under the exchange of $1\leftrightarrow 2$ and/or
$3\leftrightarrow 4$. In terms
of the spinor-free amplitudes, this implies
\begin{equation}
\begin{split}
f_{\lambda_2 \lambda_1 \lambda_3 \lambda_4}(s,t) = f_{\lambda_1
  \lambda_2 \lambda_3 \lambda_4}(s,u)\,, \\ f_{\lambda_1 \lambda_2
  \lambda_4 \lambda_3}(s,t) = f_{\lambda_1 \lambda_2 \lambda_3
  \lambda_4}(s,u)\,,
\end{split}
\label{eq:bosesymm}
\end{equation}
with $u=-s-t$. These relations provide non-trivial checks for our results.

\section{Details of the calculation}
\label{sec:comp}

The spinor-free helicity amplitudes $f_{\lambda_1 \lambda_2 \lambda_3 \lambda_4}$
can be computed as perturbative series in the QCD coupling constant
$\alpha_s$.
For a generic helicity configuration we introduce the
short hand $\lambdavec = (\lambda_1, \lambda_2, \lambda_3, \lambda_4)$ and
write
\begin{equation} \label{eq:fbare}
f_\lambdavec= \sum_{L=1}^{3} \left(\frac{\alpha_{s,b}}{2\pi}\right)^L
\,f_\lambdavec^{(L,b)} + \mathcal{O}(\alpha_{s,b}^4),
\end{equation}
where $\alpha_{s,b}$ is the \emph{bare} strong coupling constant and
$f_\lambdavec^{(L,b)}$ is the \emph{bare} perturbative
coefficient of the helicity amplitude.
Since the leading order contribution $f_\lambdavec^{(1,b)}$
to the production of two photons in gluon fusion already involves
one-loop integrals, the next-to-next-to-leading order contribution
$f_\lambdavec^{(3,b)}$ involves three-loop integrals.
The main goal of this paper is to calculate $f_\lambdavec^{(3,b)}$.

As explained in sec.~\ref{sec:hel}, we can obtain the helicity amplitudes by computing the
$\oF_i$, $i=1,...,8$ form factors. In principle, this can be achieved
straightforwardly by applying the projectors $\mathcal P_i$, $i=1,\dots,8$,
of sec.~\ref{sec:hel} to the sum of all the relevant Feynman diagrams.
At three loops, this leads to a sum of terms of the form
\begin{equation}
 \int \left(\prod_{i=1}^3 \mathcal{D}^d k_i\right)
\frac{\mathcal N(d;\{p_i\cdot p_j\},\{p_i\cdot k_j\},\{k_i \cdot
  k_j\})}{D_1^{n_1} \dots D_{10}^{n_{10}}}\,,  
\label{eq:genintnum}
\end{equation}
where $k_i$, $i=1,2,3$, are the loop momenta, $D_i$ are the propagators
of the graphs and $n_i$ are non-negative integers.
Following previous work~\cite{Henn:2020lye,Caola:2020dfu}, the integration measure for every loop
is defined as
\begin{equation}
    \int \mathcal{D}^d k_i = e^{\epsilon \gamma_E}\int \frac{d^d k_i}{i \pi^{d/2}}\,.
\end{equation}
It is convenient to treat propagators and
scalar products involving the loop momenta
on the same footings. We
do this by writing scalar products in the numerator as
additional propagators raised to negative powers.
For our problem, there are 6 scalar
products of the form $k_i\cdot k_j$ and 9 of the form $k_i\cdot p_j$,
so we can write a generic Feynman integral of the form eq.~(\ref{eq:genintnum})
  as
\begin{equation}
\int \left(\prod_{i=1}^3 \mathcal{D}^d k_i \right)
\frac{f(d;\{p_i\cdot p_j\})}{D_1^{n_1} \dots D_{15}^{n_{15}}}\,,
\label{eq:genint}
\end{equation}
where now $n_i$ can also be negative integers. We refer to each set of
inequivalent $\{D_1,...,D_{15}\}$ as an ``integral family''. Within
each family, it is well known that not all the integrals are linearly
independent. Indeed, Feynman integrals satisfy 
integration-by-parts (IBP) identities~\cite{Chetyrkin:1981qh} of
the form
\begin{align}
\int \left(\prod_{i=1}^3 \mathcal{D}^d k_i \right)
\frac{\partial}{\partial k^\mu_j}\frac{v^\mu_j}{D_1^{n_1} ... D_{m}^{n_{m}}} = 0\,,
\label{eq:IBP}
\end{align}
where $v_j$ can be any loop or external momentum. In principle, it is
possible to use these identities to 
express all the $\oF_i$ form factors in terms of a minimal set of
independent ``master integrals'' (MI)~\cite{Laporta:2000dsw}.
While all the steps described above are well-understood in principle,
 the complexity involved in intermediate stages
grows very quickly with the number
of loops and external scales. In our case, the three-loop calculation
involves 3 different families, each of which can contribute with 6 independent
crossings of the external legs, and more than $4\times 10^6$ integrals to the amplitude.
Moreover, using \eqref{eq:IBP} directly would lead to a very large number of equations
involving also many additional auxiliary integrals.
We now describe the 
procedure that we have adopted to keep the degree of complexity
manageable.

\begin{figure}[t]
 	\centering
 	\begin{subfigure}[b]{0.25\textwidth}
 		\includegraphics[width=\textwidth]{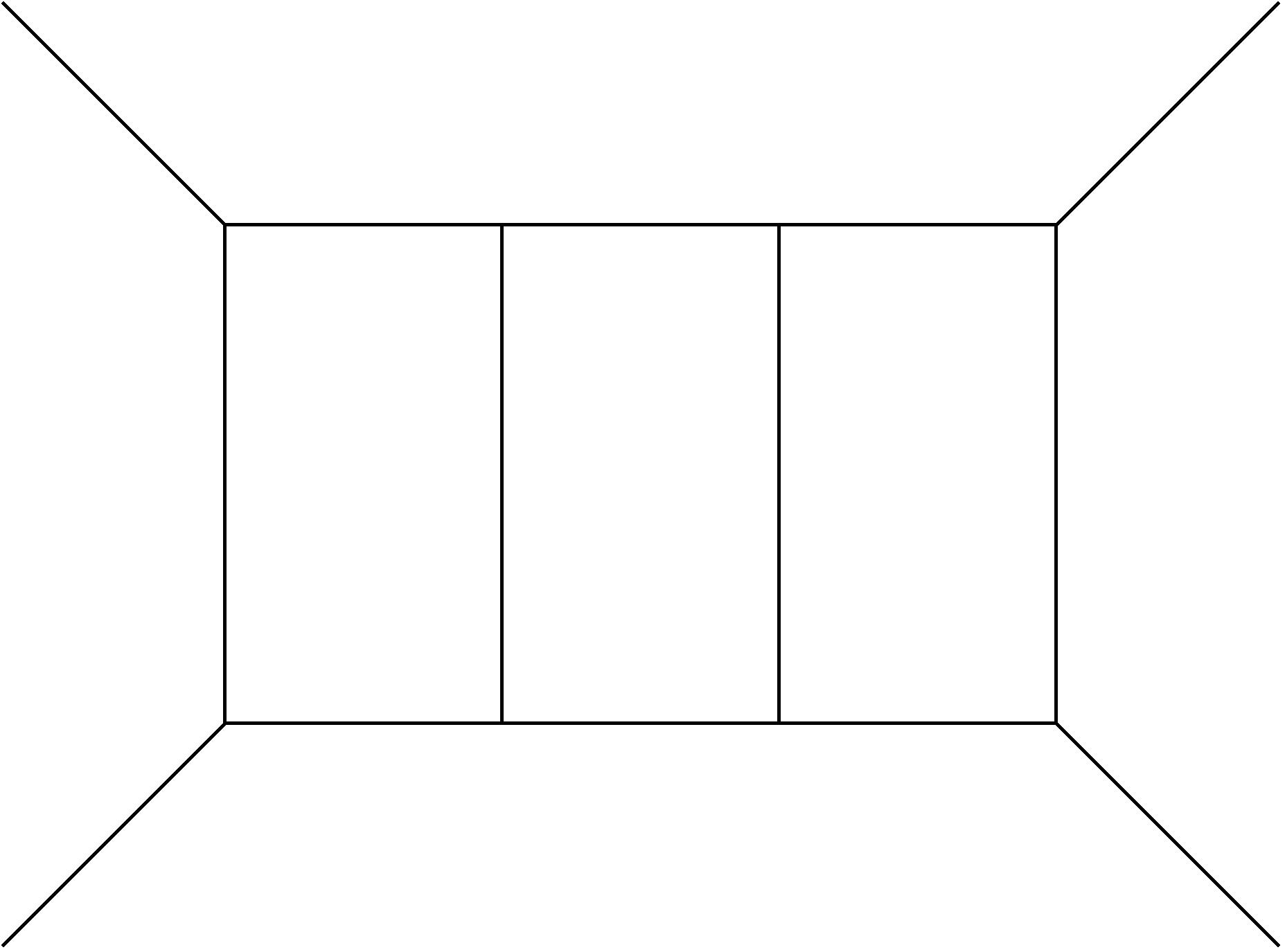}
 		\caption*{(PL)}
 	\end{subfigure}
 	\hspace*{0.05\textwidth}
 	\begin{subfigure}[b]{0.25\textwidth}
 		\includegraphics[width=\textwidth]{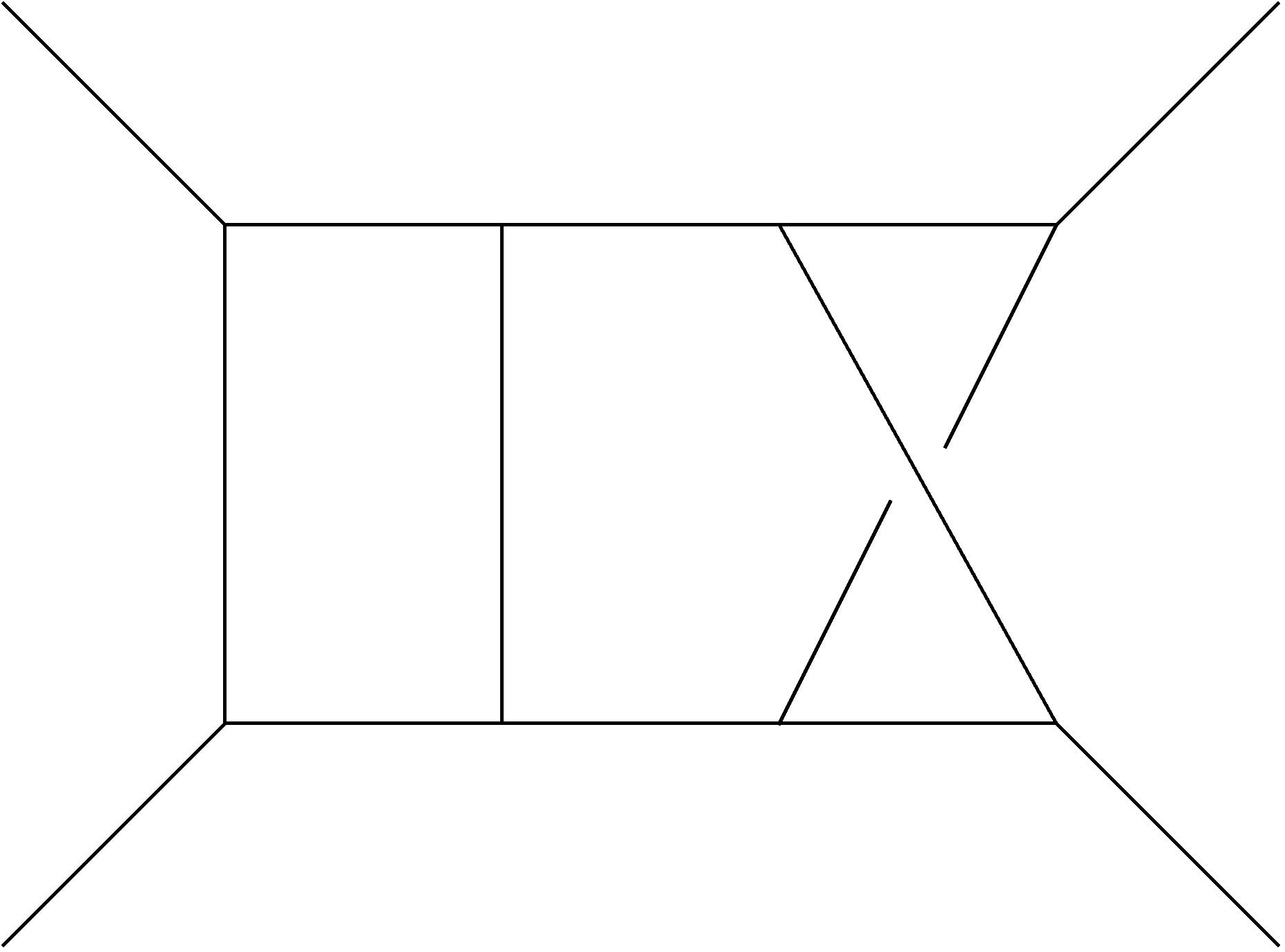}
 		\caption*{(NPL1)}
 	\end{subfigure}
 	\hspace*{0.05\textwidth}
 	\begin{subfigure}[b]{0.25\textwidth}
 		\includegraphics[width=\textwidth]{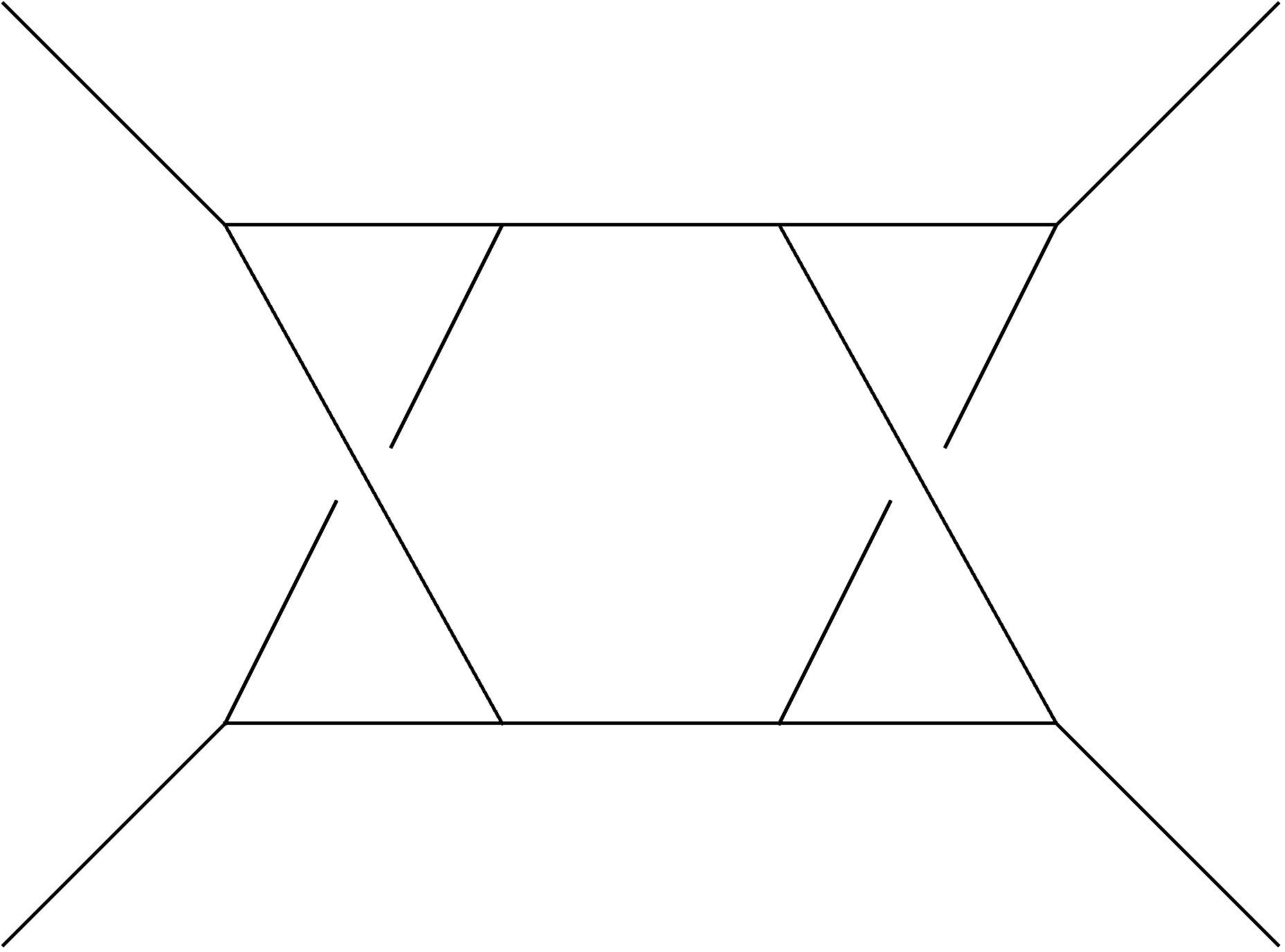}
 		\caption*{(NPL2)}
 	\end{subfigure}
 	\caption{Representative top level topologies for the planar (PL), single nonplanar (NPL1), and double nonplanar (NPL2) integral families.}
 	\label{fig:FD}
 \end{figure}

First, we generated all Feynman diagrams with
\texttt{Qgraf}~\cite{Nogueira:1991ex} and mapped each diagram to an integral family using \texttt{Reduze 2}~\cite{vonManteuffel:2012np,Studerus:2009ye} to generate the required shifts of loop momenta.
At this stage, it is useful to group  diagrams that present
similar structures together and
perform the $\mathcal P_{1,...,8}$ projections for each of these groups 
 separately. This can be done by keeping together diagrams that can be mapped to the same crossing of the same integral families. This allows us to 
 reduce redundancy in the algebraic manipulations required.  
 Examples of top sectors from our three families of integrals
  are depicted in Fig.~\ref{fig:FD}, while their complete definition can be found
 in the ancillary files. 
To evaluate the contributions to the form factors, we performed the colour, Lorentz and Dirac
algebra as well as further symbolic manipulations described in the following with
\texttt{Form}~\cite{Vermaseren:2000nd}. 

We find it important to stress that by expressing the result for each
$\oF_{1,...,8}$ in terms of a minimal set of integrals under 
crossings and shift symmetries, prior to performing the actual IBP reduction, 
we saw a significant decrease in
complexity.  This is expected, as many equivalent integrals are
combined together and redundant structures are removed. After this
simplification, we used the $\oF_i$ to construct the spinor-free helicity
amplitudes, and collected the contributions to different colour structures. This
way, we arrived at a minimal set of gauge-independent building
blocks. 
We found it useful to partial fraction the rational functions with respect
to $x$ in order to reduce their complexity.

The next step is the actual IBP integral reduction. 
We did this using an in-house implementation of the Laporta
algorithm~\cite{Laporta:2000dsw},
\texttt{Finred}~\cite{vonManteuffel:2016xki}, which exploits
syzygy-based
techniques~\cite{Gluza:2010ws,Ita:2015tya,Larsen:2015ped,Bohm:2017qme,Schabinger:2011dz,Agarwal:2020dye}
and finite-field
arithmetics~\cite{vonManteuffel:2014ixa,Peraro:2016wsq,Peraro:2019svx,vonManteuffel:2016xki}.
We found in this way that the three loop helicity amplitudes can be
expressed in terms of the 221 MIs computed in ref.~\cite{Henn:2020lye}
and crossed versions of them, for a total of $486$ MIs.\footnote{We note
  that while the reductions of some integrals were already known from
  earlier calculations~\cite{Caola:2020dfu,Caola:2021rqz}, for this process
  we had to reduce a significant number of new integrals 
  compared to those references.} We stress that these MIs are pure
functions, i.e.\ they do not have any non-trivial rational functions of $x$ or $d$
as prefactors.  Before inserting the IBP relations into the amplitude,
we partial fractioned them with respect to both $d$ and $x$. We found
that this step is crucial to keep the complexity under
control. Finally, we performed one last partial fraction decomposition of the full
amplitude after we wrote it in terms of MIs.

As a last step, we expand in $\epsilon$ and substitute the analytic results for the
MIs. All of the integrals required for our calculation
were computed in ref.~\cite{Henn:2020lye}.
Their $\epsilon$ expansion can be written in terms of Harmonic Polylogarithms (HPL), that we
define iteratively as\footnote{Note that we use the GPL notation
  of ref.~\cite{Vollinga:2004sn}, rather than the original HPL notation of
  ref.~\cite{Gehrmann:2001pz}.}
\begin{equation}
  G(\underbrace{0,\dots,0}_{n~\text{times}};x) \equiv \frac{\ln^n x }{n!},~~~~~~~
G(a_n,...,a_1;x) = \int_0^x \frac{dz}{z-a_n} G(a_{n-1},...,a_1;z),
\label{eq:HPL}
\end{equation}
with $G(x)=1$ and $a_i\in\{0,1\}$. For our case, we need to consider
polylogarithms up to weight 6, i.e.\ $n=6$ in eq.~(\ref{eq:HPL}).
We used the Mathematica package \texttt{PolyLogTools}~\cite{Duhr:2019tlz} 
to manipulate HPLs up to weight 5, augmented by a straightforward generalisation
of its routines up to the required weight 6, as well as an independent package
for multiple polylogarithms written by one of us.
As expected from the fact that there are fewer weight $\leq 6$ HPLs than MIs,
we observed a noticeable decrease in complexity for the amplitude
after we expressed it in term of HPLs. We summarise the degree of
complexity of the various steps discussed above in Tab.~\ref{tab:stats}.
\begin{table}[t]
	\centering
	\begin{tabular}{ p{9.5cm}||p{1.cm}|p{1.cm}|p{1.5cm} }
		& 1L & 2L & 3L \\
		\hline
		\hline
		Number of diagrams & 6 & 138 & 3299 \\
		\hline
		Number of inequivalent integral families & 1 & 2 & 3 \\
		\hline
		Number of integrals before IBPs and symmetries & 209 & 20935 & 4370070 \\
		\hline
		Number of master integrals & 6 & 39 & 486 \\
		\hline
		\hline
		Size of the \texttt{Qgraf} result [kB] & 4 & 90 & 2820 \\
        \hline
		Size of the \texttt{Form} result before IBPs and symmetries [kB] & 276 & 54364 & 19734644 \\
		\hline
		Size of helicity amplitudes written in terms of MIs [kB] & 12 & 562 & 304409 \\
		\hline
		Size of helicity amplitudes written in terms of HPLs [kB] & 136 & 380 & 1195 \\
	\end{tabular}
	\caption{Complexity of the various stages
		of the calculation at different loop orders.}
	\label{tab:stats}
\end{table}

Before presenting our results, we note that although the MIs have been
computed in ref.~\cite{Henn:2020lye}, for this calculation we have
decided to recompute them as an independent check. We used the same
definitions for the MIs as ref.~\cite{Henn:2020lye}, and followed the same
strategy outlined in that reference for obtaining their analytic.  First, since the basis~\cite{Henn:2020lye}
is pure and of uniform weight~\cite{Henn:2013pwa} the MIs obey  very simple differential
equations
\begin{equation}
  \mathrm{d}\vec M(\ep;s,t,u) = \ep\left[A_s\,\mathrm{d}\log(s) + A_t\,\mathrm{d}\log(t) + A_u\,\mathrm{d}\log(u)\right] \vec M(\ep;s,t,u),
  \label{eq:j}
\end{equation}
where $\vec M$ is a vector whose components are the MIs and $A_i$ are
constant matrices. Using the basis of ref.~\cite{Henn:2020lye} we
have rederived the differential equation from scratch and found
agreement. Given the simple form of eq.~\eqref{eq:j}, it is straightforward to iteratively solve
it order by order in $\ep$, modulo boundary conditions.
The only non-trivial issue is how to fix the latter. Very
interestingly, the authors of ref.~\cite{Henn:2020lye} noted that at
three loops it is enough to impose regularity conditions to fix all
boundary conditions, apart from one simple overall normalisation. The
main idea is to look at the differential equation near singular points
$s\to 0$, $t\to 0$, $u\to 0$.  Let us consider $s\to 0$ as an example.
In this limit the general solution of eq.~(\ref{eq:j}) behaves like
\begin{equation}
  \vec M \sim s^{A_s\ep} \vec M_{0,s},
\end{equation}
where $\vec M_{0,s}$ is a constant vector.
It was argued in ref.~\cite{Henn:2020lye} that the MIs considered here
can only develop branch cuts of the form $s^{-\alpha \ep}$ with $\alpha>0$.
This implies that the coefficient of $s^{\alpha\ep}$ in $s^{A_s\ep} \vec M_{0,s}$ must vanish for $\alpha>0$.
As a consequence, there must exist non-trivial relations between
different MIs in the $s\to 0$ limit. When combined with analogous relations
derived from the limits $t,u\to 0$, the authors of ref.~\cite{Henn:2020lye} found that for the case under study one can completely constrain all the
boundary conditions up to an overall normalisation factor. 
We
have independently verified that this is the case, which allowed
us to rederive an analytic expression for all the master integrals. 
We have then verified that our results to weight 6 are identical
to the ones of ref.~\cite{Henn:2020lye}, provided that the latter are analytically continued
to the physical Riemann sheet. 
Since in ref.~\cite{Henn:2020lye} final results are only
presented for one single crossing, 
for convenience we decided to provide analytic results for all the three-loop master integrals and all their crossings in the ancillary files accompanying this publication. 
We also provide weight 6 results for a uniform-weight basis of the two-loop integrals.

\section{UV renormalisation and IR regularisation}
\label{sec:uvir}

Following the steps outlined above, we obtained analytical expressions
for the bare helicity amplitudes $f_\lambdavec^{(L,b)}$ defined in
eq.~(\ref{eq:fbare}) for $L=1,2,3$.
The $f_\lambdavec^{(L,b)}$ are affected by both ultraviolet
(UV) and infrared (IR) divergences, which manifest themselves as poles in the
dimensional regularisation
parameter $\epsilon = (4-d)/2$.
While the former are removed by UV renormalisation,
the latter can be regularised using universal IR operators 
acting on lower-loop amplitudes. We now discuss in detail how this
can be done.

We first consider UV divergences.  We define $\alpha_s(\mu)$ to be the
renormalised strong coupling constant in the $\overline{ \rm MS}$
scheme at the scale $\mu$
\begin{equation}
  S_{\epsilon}\mu_0^{2\ep}\alpha_{s,b} = \mu^{2\ep} \alpha_s(\mu) Z[{\alpha_s(\mu)}],
  \label{eq:asren}
\end{equation}
with
$S_{\epsilon} = (4\pi)^{\epsilon}e^{-\gamma_E \epsilon}$
and
\begin{equation}
Z[\alpha] = 1-\frac{\beta_0}{\ep}\left(\frac{\alpha_s}{2\pi}\right) + 
\left(\frac{\beta_0^2}{\epsilon^2}-\frac{\beta_1}{2\epsilon}\right)
\left(\frac{\alpha_s}{2\pi}\right)^2+ \mathcal O(\alpha_s^3).
\label{eq:renorm}
\end{equation}
The first two coefficients of the QCD beta function read
\begin{equation}
\beta_0 = \frac{11}{6}C_A - \frac{2}{3}T_F n_f\,,\qquad 
\beta_{1} = \frac{17}{6}C_A^2 -  T_F n_f \left(\frac{5}{3}C_A+C_F\right).
\end{equation}
We then expand the spinor-free helicity amplitudes  $f_\lambdavec$ in terms of the
renormalised strong coupling $\alpha_s(\mu)$ as
\begin{equation} \label{eq:fUV}
f_\lambdavec = \sum_{L=1}^{3} \left(\frac{\alpha_{s}(\mu)}{2\pi}\right)^L
f_\lambdavec^{(L)}.
\end{equation}
The expression for the renormalized amplitudes $f_\lambdavec^{(L)}$ can be obtained by
substituting eq.~(\ref{eq:asren}) in eq.~\eqref{eq:fbare} and expanding
in the renormalised coupling. 
For convenience, we will set $\mu^2=s$ in the following. The result for
arbitrary scale can be easily obtained using renormalisation group
methods.

We now consider IR divergences.  The IR structure of the amplitude is
governed by the soft and collinear behaviour of
virtual quarks and gluons and it is universal, i.e. it only depends on
the colour and nature of the external legs.
This allows one to write the renormalised amplitude as
\begin{align}
 f_\lambdavec^{(1)} &= f_\lambdavec^{(1,{\rm fin})}, \notag\\  f_\lambdavec^{(2)} &=
  \mathcal I_1\, f_\lambdavec^{(1)} + f_\lambdavec^{(2,{\rm fin})}, \notag\\ 
  f_\lambdavec^{(3)} &= \mathcal I_2\, f_\lambdavec^{(1)} + \mathcal
  I_1\,f_\lambdavec^{(2)} + f_\lambdavec^{(3,{\rm fin})},
\label{eq:catani}
\end{align}
where $f_\lambdavec^{(i,\rm fin)}$ are finite in four dimensions. The IR
structure is encoded in the operators $\mathcal I_i$, that for our
case (with $\mu^2=s$) read~\cite{Catani:1998bh}
\begin{align}
\mathcal I_1(\epsilon) &= -\frac{e^{i\pi\epsilon}e^{\gamma_E\epsilon}}{\Gamma(1-\epsilon)}
\left(\frac{C_A}{\epsilon^2}+\frac{\beta_0}{\epsilon}\right),
\notag\\
\mathcal I_2(\epsilon) &= -\frac{1}{2}\mathcal I_1(\epsilon)
\left(\mathcal I_1(\epsilon)+\frac{2\beta_0}{\epsilon}\right)+
\frac{e^{-\gamma_E\epsilon}\Gamma(1-2\epsilon)}{\Gamma(1-\epsilon)}
\left(\frac{\beta_0}{\epsilon}+K\right)\mathcal I_1(2\epsilon) + 2\frac{e^{\epsilon\gamma_E}}{\Gamma(1-\epsilon)}H_g\,,
\label{eq:poles}
\end{align}
where $K$ is the next-to-leading-order coefficient of the cusp anomalous dimension
\begin{equation}
K = \left(\frac{67}{18}-\frac{\pi ^2}{6}\right)C_A - \frac{10}{9}n_f
  T_F,  
\label{eq:cusp}
\end{equation}
and~\cite{Harlander:2000mg}
\begin{equation}
H_g = \frac{1}{2\epsilon} \left[ \left(\frac{\zeta
    (3)}{4}+\frac{5}{24}+\frac{11 \pi ^2}{288}\right) C_A^2 +T_F n_f
  \left(\frac{C_F}{2}-\left(\frac{29}{27}+\frac{\pi^2}{72}\right)C_A\right)+\frac{10
    }{27}T_F^2 n_f^2 \right]\,.
\label{eq:Hterm}
\end{equation}
In eqs~\eqref{eq:catani} we used the fact that diphoton production in
gluon fusion starts at one loop.  The finite remainders for the
helicity amplitudes $f_\lambdavec^{(L,fin)}$ up to three loops are the
main result of this paper, and we provide analytic results for them in the
ancillary files.

\section{Checks and structure of the result}
\label{subsec:res}

We have performed various checks on the correctness of our results.
First, we have employed two derivations of the three-loop
$\mathcal F_i$ form factors at the integrand level and verified
that they agree.
We have also compared the one- and two-loop helicity
amplitudes against the results of ref.~\cite{Bern:2001df}
and found agreement.
To validate our numerical evaluation procedure, we also checked the
helicity-summed one-loop squared amplitude against
\texttt{OpenLoops}~\cite{Cascioli:2011va, Buccioni:2019sur}, and one helicity configuration
at two loops against \texttt{MCFM}~\cite{Campbell:2011bn,Boughezal:2016wmq}.
Finally, we have verified that the UV and IR poles up to three loops
follow the structure described in the previous section.  This provides a
strong check of the correctness of the three-loop amplitudes.

We now discuss the general structure of our result.
The amplitude can be expressed in terms of the two quadratic Casimirs
$C_A$ and $C_F$ and the flavour structures $n_f$, $n_f^V$ and $n_f^{V_2}$
defined in eq.~\eqref{eq:nfdef}.
At $L$ loops the amplitude is a
homogeneous degree-$L$ polynomial in these 5 variables. At one-loop,
the amplitude is only proportional to $n_f^{V_2}$, since the two
photons must both couple to the same fermion line. At two loops, the
structures $n_f^{V_2}\times \{C_F,C_A\}$ appear in the bare amplitude.
The finite remainder contains in addition a term proportional to
$n_f^{V_2} n_f$ stemming from $\beta_0$ in
the UV/IR regularisation.  We note that there is no $(n_f^V)^2$ contribution. It is easy to understand why this is the case. The $(n_f^V)^2$ colour factor
only appears if the two photons are attached to two different (closed)
fermion lines.
Such diagrams do appear at two loops, but they are of the form of two $\gamma g g^*$ one-loop
triangles connected through a gluon propagator.
Due to an argument analogous to Furry's theorem, these diagrams give
no net contribution to the amplitude.
A similar argument allows one to conclude that there
is no net contribution from Feynman diagrams with colour factors $n_f
(n_f^{V})^2$ at three loops. Furthermore, it is easy to see that 
the structure $n_f^2 n_f^{V_2}$ is absent in the three-loop bare
amplitude.\footnote{We note however that the $n_f^2 n_f^{V_2}$
  structure contributes to our finite remainders, since it is induced by 
  the $n_f$ dependence of the UV/IR counterterms.}
Since there is no $(n_f^V)^2$ contribution at lower loops, the $(n_f^V)^2$
term in the bare three loop amplitude must be finite.
We observe, however, that it is non-zero. Indeed, at three loops this colour factor appears in
triple-box diagrams for which the Furry argument outlined above is no longer
applicable. 

We now move to the discussion of the kinematic features of the three-loop
amplitude, i.e.\ its $x$ dependence. The amplitude contains terms of
the form $G(a_1,...,a_n;x)/x^k$ ($-2\leq k \leq 2$)
and $G(a_1,...,a_n;x)/(1-x)^k$ ($1\leq k \leq 2$)
, 
where $a_i\in\{0,1\}$, $0\le n \le 6$, and $G$ are the
Harmonic Polylogarithms defined in eq.~\eqref{eq:HPL}. 
Instead of the HPLs, we found it useful to also consider the alternative functional basis described in ref.~\cite{Caola:2020dfu} to speed up the numerical evaluation of the final result.
Using the algorithm of~\cite{Duhr:2011zq}, we have constructed a basis of logarithms, classical polylogarithms and multiple polylogarithms to rewrite the HPLs without introducing any new spurious singularities.
We used products of lower weight functions whenever possible and preferred functions whose series representation requires a small number of nested sums.
In this way, we found that 23 independent transcendental functions and products thereof suffice to represent our HPLs up to weight 6.
The new basis consists of 2 logarithms, $\ln(x)$
$\ln(1-x)$, 12 classical polylogarithms, $\Li_2$ of $x$,
$\Li_3$ of $x$ and $1-x$ and $\Li_4$, $\Li_5$, $\Li_6$ of $x, 1-x$ and
$-x/(1-x)$, as well as 9 multiple polylogarithms
$\Li_{3,2}(1,x),\Li_{3,2}(1-x,1),\Li_{3,2}(x,1),\Li_{3,3}(1-x,1),\Li_{3,3}(x,1),\Li_{3,3}\left(\frac{-x}{1-x},1\right),\Li_{4,2}(1-x,1),\Li_{4,2}(x,1),\Li_{2,2,2}(x,1,1)$.
Here, we follow the conventions of ref.~\cite{Vollinga:2004sn} and
define
\begin{equation}
\label{eq:Linm}
\Li_{m_1,...,m_k}(x_1,...,x_k) = \sum_{i_1>...>i_k>0} \frac{x_1^{i_1}}{i_1^{m_1}} \, ... \, \frac{x_k^{i_k}}{i_k^{m_k}} \,.
\end{equation}
In the ancillary files, we provide our analytic results written both in terms of HPLs and in terms of this minimal set of functions.
For convenience, we 
also provide results for the  finite remainders of the one- and two-loop helicity amplitudes 
up to weight 6.

Finally, we present our results. Although intermediate
expressions are rather complicated, see Table~\ref{tab:stats}, we find that the final
results are remarkably compact. The $\lambdavec =(++++)$ helicity configuration is
particularly simple. This is of course expected, since the one-loop
amplitude does not have support on any cut, hence it is purely rational
rather than a weight-2 function. This simplicity persists at higher loops.
For illustration, we now report here the result for the finite reminders
defined in eq.~\eqref{eq:catani} up to three loops for this helicity
configuration. At one and
two loops one has
\begin{align}
  f^{(1,{\rm fin})}_{++++} &= 2 n_f^{V_2},
\label{ampppppLO}\\
  f^{(2,{\rm fin})}_{++++} &= 2 n_f^{V_2} \left(2 C_A - 3 C_F + i\pi\beta_0 \right).
\label{ampppppNLO}
\intertext{At three loops, we write the finite remainder as}
f^{(3,{\rm fin})}_{++++} &= 
\Delta_1(x)\, n_f^{V_2} C_A^2
+ \Delta_2(x)\, n_f^{V_2} C_A C_F
+ \Delta_3(x)\, n_f n_f^{V_2}  C_A
+ \Delta_4(x)\, (n_f^V)^2 C_A \nonumber \\
&\quad + \Delta_5(x)\, n_f^{V_2} C_F^2 
+ \Delta_6(x)\, (n_f^V)^2 C_F
+ \Delta_7(x)\, n_f n_f^{V_2} C_F
+ \Delta_8(x)\, n_f^2 n_f^{V_2} \nonumber \\
&\quad + \{(x)\leftrightarrow(1-x)\}
\label{amppppp}
\end{align}
with
\allowdisplaybreaks
\begin{align}
\Delta_1(x) &=
-
\mfrac{23 L_1 (L_1+2 i \pi )}{9 x^2}+
\mfrac{32 L_1 (L_1+2 i \pi )-
46 (L_1+i \pi )}{9x}-
\mfrac{17}{36} L_0^2-
\mfrac{19}{36} L_0 L_1+
\mfrac{1}{9}L_0-2 i \pi L_0
\nonumber \\ & \quad
+\mfrac{1}{288}\pi ^4
-\mfrac{373}{72} \zeta_3
-\mfrac{185}{72} \pi ^2
+\mfrac{4519}{324}
+\mfrac{1}{2}i \pi  \zeta_3
+\mfrac{11}{144} i \pi ^3
+\mfrac{157}{12} i \pi 
+ \mfrac{43}{9} L_0 x
\nonumber \\ & \quad
-\mfrac{7}{9} x^2 \left((L_0-L_1)^2
+\pi ^2\right)
\,, \nonumber \\
\Delta_2(x) &=
\mfrac{8 L_1 (L_1+2 i \pi )}{3 x^2}
+\mfrac{16 (L_1+i \pi )
-{8} L_1 (L_1+2 i \pi )}{3x}
-\mfrac{1}{3}L_0^2
+\mfrac{5 }{6}L_0 L_1
+\mfrac{17}{3}L_0+i \pi  L_0
-\mfrac{5 }{12}\pi ^2
\nonumber \\ & \quad
-\mfrac{199}{6}
-{8} i \pi 
-\mfrac{16}{3} L_0 x
+\mfrac{4}{3} x^2 \left((L_0-L_1)^2+\pi ^2\right)
\,, \nonumber \\
\Delta_3(x) &= 
\mfrac{L_1 (L_1+2 i \pi )}{18 x^2}
+\mfrac{2(L_1+i \pi )
- L_1 (L_1+2 i \pi )}{18x}
-\mfrac{1}{36}L_0^2
+\mfrac{1}{36}L_0 L_1
-\mfrac{1}{9}L_0
-\mfrac{61 }{36}\zeta_3
+\mfrac{475}{432} \pi ^2
\nonumber \\ & \quad
-\mfrac{925}{324}
-\mfrac{1}{72}i \pi ^3
-\mfrac{175 }{54}i \pi 
+\mfrac{2}{9} L_0 x
+\mfrac{1}{36} x^2 \left((L_0-L_1)^2+\pi ^2\right)
\,, \nonumber \\
\Delta_4(x) &=
-\mfrac{5 L_1 (L_1+2 i \pi )}{4 x^2}
+\mfrac{ L_1 (L_1+2 i \pi )-8 (L_1+i \pi )}{2x}
+\mfrac{1}{4}L_0^2
-\mfrac{1}{4}L_0 L_1
-{2} L_0
-{6} \zeta_3
+\mfrac{1}{8}\pi ^2
-\mfrac{1}{2}
\nonumber \\ & \quad 
+{4} L_0 x
-x^2 \left((L_0-L_1)^2+\pi ^2\right)
\,, \nonumber \\
\Delta_5(x) &= 
-\mfrac{L_1 (L_1+2 i \pi )}{x^2}
+\mfrac{L_1 (L_1+2 i \pi )-2 (L_1+i \pi )}{x}
-\mfrac{1}{2}L_0^2
-i \pi  L_0
+\mfrac{39}{4}
+i \pi 
+{2} L_0 x
\nonumber \\ & \quad
-\mfrac{1}{2} x^2 \left((L_0-L_1)^2+\pi ^2\right)
\,, \nonumber \\
\Delta_6(x) &= 
\mfrac{10 L_1 (L_1+2 i \pi )}{3 x^2}
+\mfrac{32 (L_1+i \pi )
-{4} L_1 (L_1+2 i \pi )}{3x}
-\mfrac{2}{3} L_0^2
+\mfrac{2}{3} L_0 L_1
+\mfrac{16}{3} L_0
+{16} \zeta_3
-\mfrac{1}{3}\pi ^2
\nonumber \\ & \quad
+\mfrac{4}{3}
-\mfrac{32}{3} L_0 x
+\mfrac{8}{3} x^2 \left((L_0-L_1)^2+\pi ^2\right)
\,, \nonumber \\
\Delta_7(x) &= 
\mfrac{5 L_1 (L_1+2 i \pi )}{3 x^2}
+\mfrac{10 (L_1+i \pi )
-8 L_1 (L_1+2 i \pi )}{3x}
+\mfrac{2}{3} L_0^2
+\mfrac{1}{3}L_0 L_1
-\mfrac{10}{3}L_0+2 i \pi L_0
+{4} \zeta_3
\nonumber \\ & \quad
-\mfrac{\pi ^2}{6}
+{5}
-{3} i \pi 
-\mfrac{10 }{3}L_0 x
+\mfrac{1}{3} x^2 \left((L_0-L_1)^2+\pi ^2\right)
\,, \nonumber \\
\Delta_8(x) &= 
-\mfrac{23 }{216}\pi ^2
+\mfrac{5 }{27}i \pi 
\,.
\label{ampppppNNLO}
\end{align}
In eq.~(\ref{ampppppNNLO}), we have defined $L_0=\ln(x)$ and $L_1=\ln(1-x)$. 
We note that these are the only transcendental functions
that are needed to describe our result.

\begin{figure}[h]
	\centering
	\begin{subfigure}[b]{0.45\textwidth}
		\includegraphics[width=\textwidth]{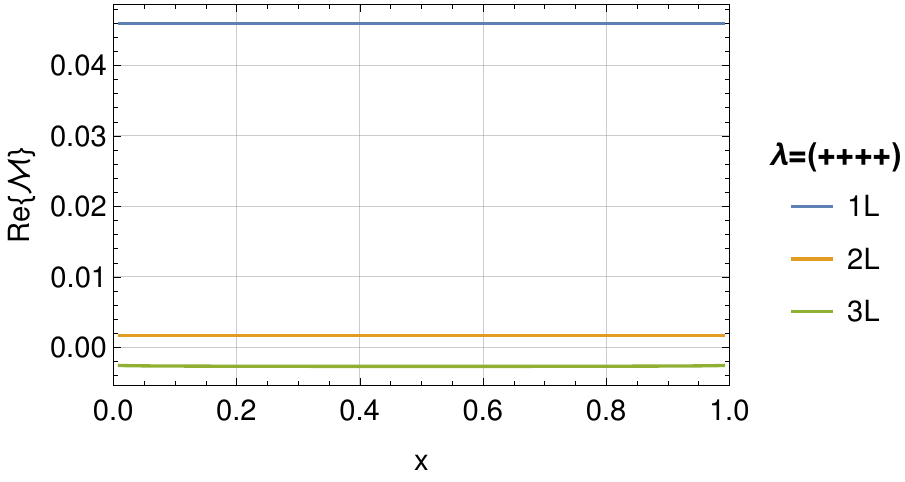}
	\end{subfigure}
	\begin{subfigure}[b]{0.40\textwidth}
		\includegraphics[width=\textwidth]{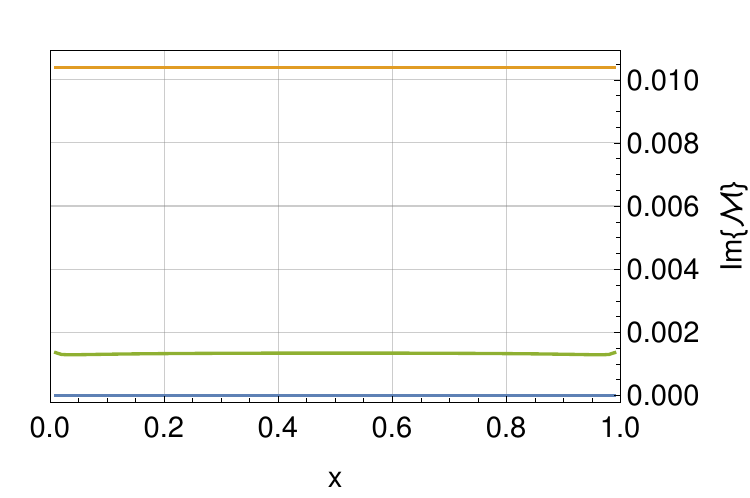}
	\end{subfigure}
	\begin{subfigure}[b]{0.45\textwidth}
		\includegraphics[width=\textwidth]{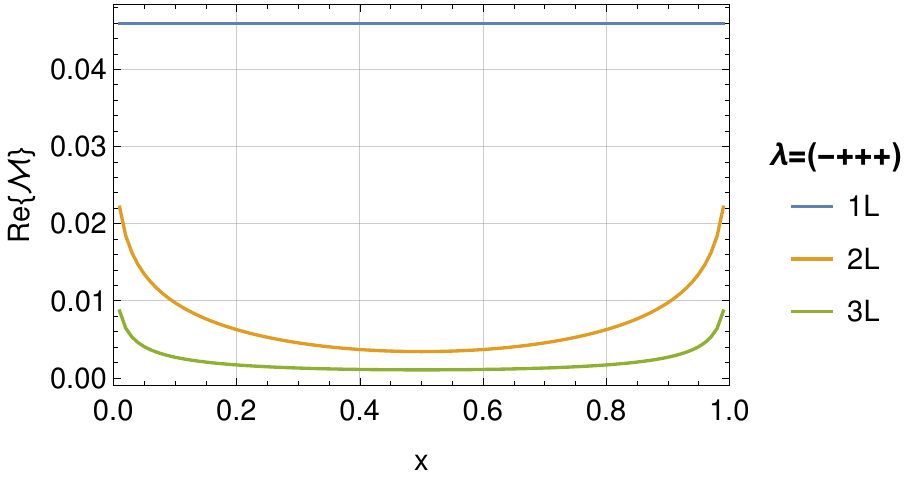}
	\end{subfigure}
	\begin{subfigure}[b]{0.41\textwidth}
		\includegraphics[width=\textwidth]{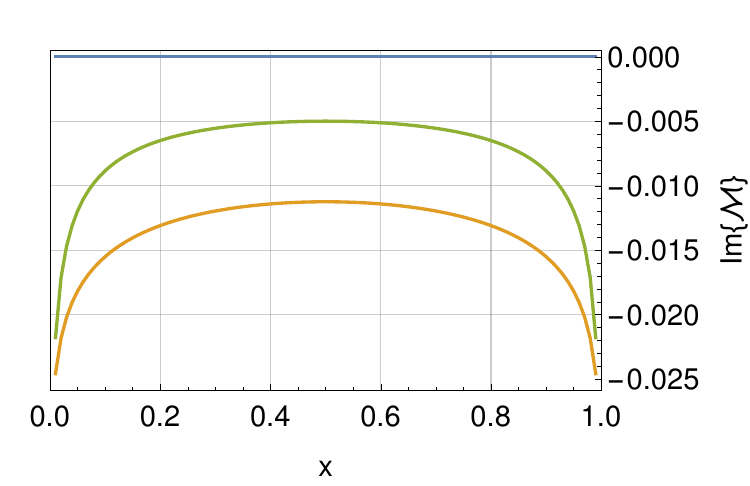}
	\end{subfigure}
	\begin{subfigure}[b]{0.45\textwidth}
		\includegraphics[width=\textwidth]{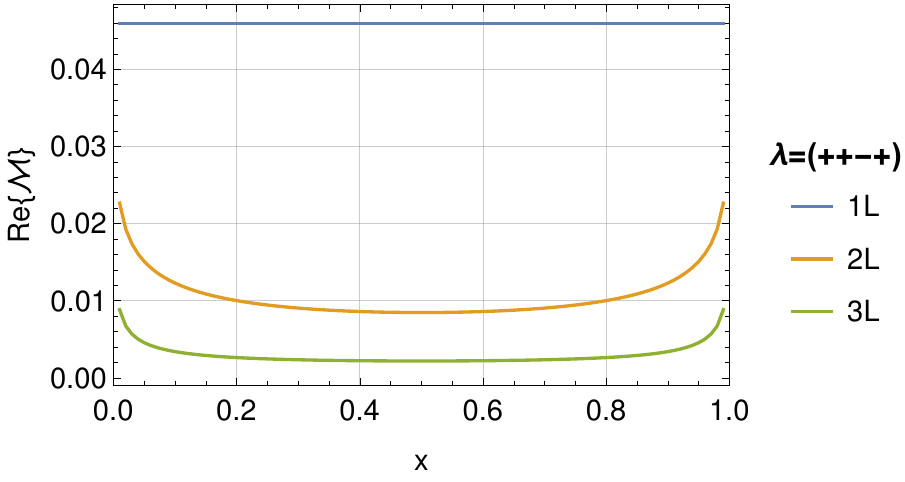}
	\end{subfigure}
	\begin{subfigure}[b]{0.41\textwidth}
		\includegraphics[width=\textwidth]{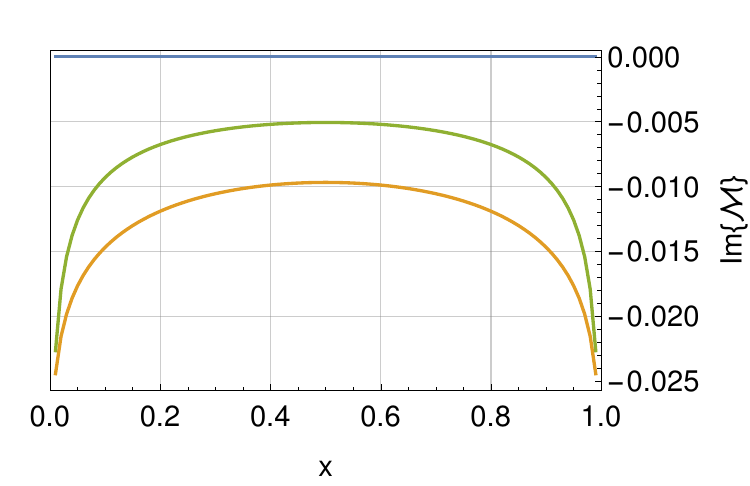}
	\end{subfigure}
	\begin{subfigure}[b]{0.46\textwidth}
		\includegraphics[width=\textwidth]{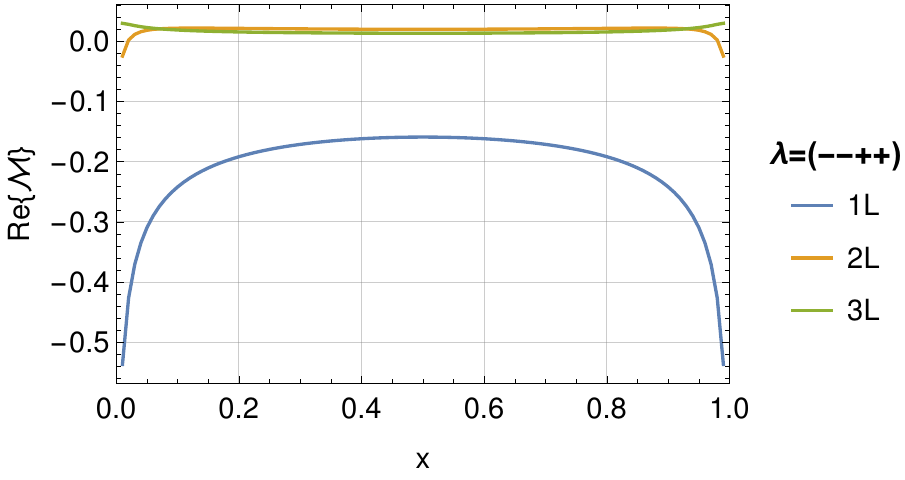}
	\end{subfigure}
	\begin{subfigure}[b]{0.40\textwidth}
		\includegraphics[width=\textwidth]{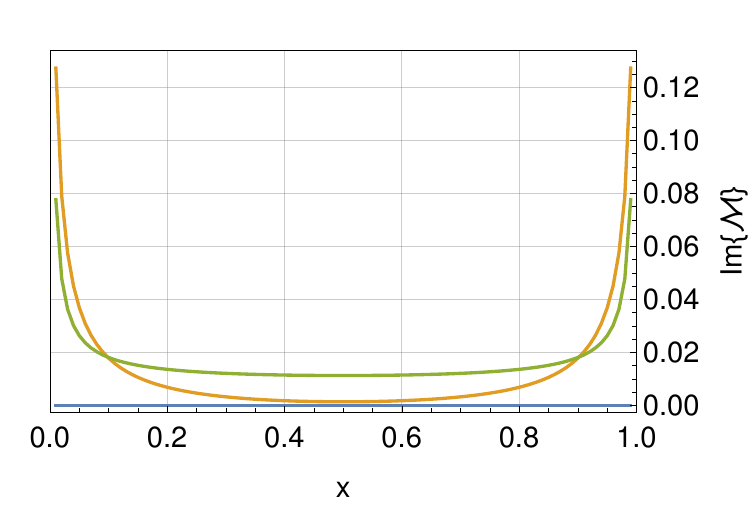}
	\end{subfigure}
	\begin{subfigure}[b]{0.47\textwidth}
		\includegraphics[width=\textwidth]{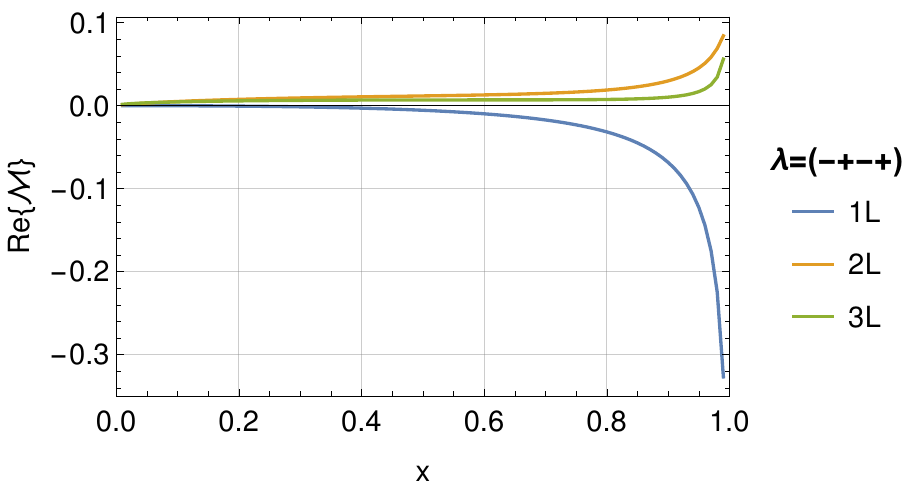}
	\end{subfigure}
	\begin{subfigure}[b]{0.40\textwidth}
		\includegraphics[width=\textwidth]{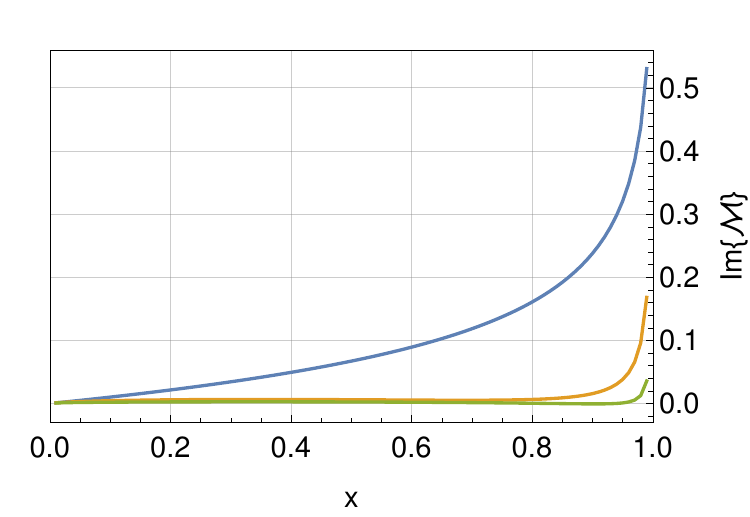}
	\end{subfigure}
	\caption{Finite remainders 
	$\mathcal{M}^{(L)}_\lambdavec \equiv \left(\frac{\alpha_s}{2\pi}\right)^L \, f^{(L,{\rm fin})}_\lambdavec$ as functions of $x=-t/s$.}
	\label{fig:numhel}
\end{figure}

Although the results for the remaining helicity configurations are still
rather compact, they are much larger than for the $\lambdavec =(++++)$
case. We provide them in electronically-readable
format, attached the arXiv submission of this paper.  In
Figure~\ref{fig:numhel}
we plot our result for the one-, two- and three-loop finite remainders
$\mathcal{M}^{(L)}_\lambdavec \equiv \left(\frac{\alpha_s}{2\pi}\right)^L f_\lambdavec^{(L,\rm fin)}$ as functions of $x$. We fix $\as = 0.118$
and show graphs for the helicity configurations $\lambdavec =(++++)$, $(-+++)$, $(++-+)$, $(--++)$ and $(-+-+)$. 
All the other helicity amplitudes can be obtained from these
through Bose symmetry ($x\leftrightarrow 1-x$) and parity.

\section{Conclusions}
\label{sec:concl}
In this paper, we have computed the
helicity amplitudes for the process $gg\to\gamma\gamma$ in three-loop massless QCD.
This is the
last missing ingredient required for the calculation of the NNLO QCD
corrections to diphoton production in the $gg$ channel.
For our analytical three-loop calculation, we have adopted a new projector-based prescription
to compute helicity amplitudes in the 't Hooft-Veltman scheme.
The expressions at the intermediate stages of our calculation were quite sizable, and
we employed recent ideas for the demanding integration-by-parts reductions.
Our final results though are remarkably compact. They can be expressed either in
terms of standard Harmonic Polylogarithms of weight up to six, or in terms of only 23
transcendental functions defined by up to three-fold sums.
This makes the numerical
evaluation of our result both fast and numerically stable. Analytical
results for both choices of the transcendental functions are provided
in the ancillary files that accompany this publication. 

We envision several
possible future directions of investigations. On a more phenomenological
side, it would be interesting to combine our results with those
of refs~\cite{Badger:2021imn,Badger:2021ohm} to obtain NNLO predictions
for the $gg\to\gamma\gamma$ process. On a more theoretical side, the
simplicity of our final results begs for an exploration of new ways to
perform multiloop calculations. Finally, it would be very interesting
to promote our calculation to the fully non-abelian case and consider
three-loop scattering amplitudes for the $gg\to gg$ process. We look
forward to pursuing these lines of investigation in the future.

\section*{Acknowledgements}
The research of PB and FC was supported by the ERC
Starting Grant 804394 \textsc{HipQCD} and by the UK Science and Technology Facilities Council (STFC) under grant ST/T000864/1.
AvM was supported in
part by the National Science Foundation through Grant
2013859.
LT was supported by the Excellence Cluster \textsc{ORIGINS} funded by the
Deutsche Forschungsgemeinschaft (DFG, German Research Foundation) under Germany's Excellence Strategy - EXC-2094 - 390783311,
by the ERC Starting Grant 949279 \textsc{HighPHun}
and by the Royal Society grant URF/R1/191125.
Feynman graphs were drawn with \texttt{Jaxodraw}~\cite{Binosi:2003yf,Vermaseren:1994je}.

\bibliographystyle{JHEP}
\bibliography{ggaa}

\end{document}